# A Note on an R$^2$ Measure for Fixed Effects in the Generalized Linear Mixed Model

Lloyd J. Edwards[a]

[a]Department of Biostatistics, 3105H McGavran-Greenberg, CB# 7420, University of North Carolina, Chapel Hill, NC 27599, U.S.A.

email: Lloyd_Edwards@unc.edu

**Abstract**

Using the LRT statistic, a model $R^2$ is proposed for the generalized linear mixed model for assessing the association between the correlated outcomes and fixed effects. The $R^2$ compares the full model to a null model with all fixed effects deleted.

## 1. Introduction

The value and familiarity of the $R^2$ statistic in the linear univariate model with normal errors creates great interest in extending it to nonlinear models and/or models with nonnormal errors. By using standard linear model theory, Edwards et al. (2008) defined an $R^2$ statistic for fixed effects in the linear mixed model based on an appropriate REML $F$ statistic, denoted $R^2_\beta$. As a result of the approximate REML $F$ statistic (Kenward and Roger 1997), $R^2_\beta$ was defined using only a single model and the statistic also generalizes to define a partial $R^2$ statistic for marginal (fixed) effects of all sorts. The choice of the null model plays a central role in defining an $R^2$ measure. $R^2_\beta$ compares the full model to a null model with all fixed effects deleted (except typically the intercept) while retaining exactly the same covariance structure. In the special case of a multivariate model that is expressed as linear mixed model and a linear mixed model hypothesis that coincides exactly with a multivariate linear hypothesis, $R^2_\beta$ reduces exactly to a multivariate measure of association using the Hotelling-Lawley-Trace statistic. The results of the research helped to provide valuable insight into $R^2$ measures in linear mixed models for the analysis of longitudinal data.

When considering models other than the standard linear regression model with iid errors, including generalized linear models, $R^2$ measures based on the likelihood ratio test (LRT) are preferable (Allen and Le 2008). In particular, use of the "log-likelihood ratio" $R^2$ for the multiple logistic regression model, denoted $R^2_L$, dates back to the early 1970's (McFadden 1974). This statistic is reported in commonly used software packages such as SAS, SPSS, and STATA.

However, very little work has been done to define an $R^2$ analog for the generalized linear mixed model. Generalized linear mixed models (GLMMs) are an extension of generalized linear models and are often used for correlated (clustered), nonnormal data. We extend the results of previous authors (McFadden 1974, Magee 1990, Menard 2000, Menard 2002, Edwards et al. 2008) to propose a model $R^2$ measure for fixed effects in the generalized linear mixed model where the focus is on longitudinal data. The proposed $R^2$ measure for the GLMM is defined using the LRT and compares the full model to a null model with all fixed effects deleted (except typically the intercept) while retaining exactly the same covariance structure. The proposed $R^2$ for the GLMM measures multivariate association between the correlated outcomes and fixed effects in the generalized linear mixed model.

## 2. The Generalized Linear Mixed Model

The generalized linear mixed model (GLMM) is perhaps the most commonly used random effects model for discrete outcomes. We summarize the basic features of the GLMM here with a focus on longitudinal data. For more details, see Breslow and Clayton (1993) and Tuerlinckx et al. (2006). With $N$ independent sampling units (often *persons* in practice) and conditionally on the random effects $\boldsymbol{u}_i$ ($m \times 1$), assume that the responses $Y_{ij}$ of $\boldsymbol{Y}_i$ ($p_i \times 1$) are independent with density function that is a member of the exponential family, i.e.,

$$f(Y_{ij}|\boldsymbol{u}_i) = \exp[\{Y_{ij}\theta_{ij} - b(\theta_{ij})\}/a(\phi) + c(Y_{ij}, \phi)] , \qquad (1)$$

for some functions $a$, $b$, and $c$. The conditional mean is $E(Y_{ij}|\boldsymbol{u}_i) = b'(\theta_{ij})$ and conditional variance is $\text{var}(Y_{ij}|\boldsymbol{u}_i) = b''(\theta_{ij})a(\phi)$. The conditional mean satisfies a linear regression model $g[E(Y_{ij}|\boldsymbol{u}_i)] = \boldsymbol{x}'_{ij}\boldsymbol{\beta} + \boldsymbol{z}'_{ij}\boldsymbol{u}_i$, where $g(\ )$ is referred to as a *link* function, $\boldsymbol{\beta}$ ($q \times 1$) is a vector of unknown fixed effect parameters, $\boldsymbol{x}_{ij}$ ($q \times 1$) and $\boldsymbol{z}_{ij}$ ($m \times 1$) are vectors of fixed and random effect explanatory variables (the first element of $\boldsymbol{x}_{ij}$ is a 1). The conditional variance can be written as $\text{var}(Y_{ij}|\boldsymbol{u}_i) = \phi a_{ij}\text{var}[E(Y_{ij}|\boldsymbol{u}_i)]$, where $a_{ij}$ is a known constant. The conditional mean also satisfies

$$E(Y_{ij}|\boldsymbol{u}_i) = h(\boldsymbol{x}'_{ij}\boldsymbol{\beta} + \boldsymbol{z}'_{ij}\boldsymbol{u}_i) , \qquad (2)$$

with $h = g^{-1}$. The random effects $\boldsymbol{u}_i$ are assumed to be sampled from a (multivariate) normal distribution with mean $\boldsymbol{0}$ ($m \times 1$) and covariance matrix $\boldsymbol{\Sigma} = \boldsymbol{\Sigma}(\boldsymbol{\alpha})$ ($m \times m$) that depends on a vector $\boldsymbol{\alpha}$ ($r \times 1$) of unknown variance components. We have thus described a GLMM with a (multivariate) normal mixing distribution for the random effects. As noted by Tuerlinckx et al. (2006), this is the model most often applied in practice.

Several methods have been proposed for inference and estimation in the GLMM. Tuerlinckx et al. (2006) reviews inference techniques for the GLMM with normal mixing distribution, discusses advantages and disadvantages of the methods, and mention software packages in which they are implemented. Tuerlinckx et al. (2006) noted three types of linear hypothesis tests of fixed effects are usually considered: LRT, a Wald test, and a score test. The Wald and score tests are approximations to the LRT and hence the p-values of the LRT are more exact (Tuerlinckx et al. 2006). However, only a single model is needed for the Wald and score tests ($R^2_\beta$ for the linear mixed model also uses only a single model) whereas both the null and full models are needed for the LRT.

## 3. A Model $R^2$ Measure for Fixed Effects in the Generalized Linear Mixed Model

The results of Edwards et al. (2008) provides the basis for many of the assumptions we use for the proposed $R^2$ measure for the GLMM. Similar to the linear mixed model, the GLMM explicitly specifies not only the mean structure, but also the covariance structure. Hence three types of model comparisons can occur. I) Compare mean models with the same covariance structure. Nested mean models are the most common. II) Compare covariance models with the same mean structure. Two GLMMs may be nested or nonnested in the covariance models. III) Compare GLMMs with different mean and different covariance structures. Consequently any definition of an $R^2$ measure for the GLMM must account for the distinction between the proportion of variation in the response explained by the fixed effects (in the mean model) and the proportion explained by the random effects (in the covariance model). The same distinction arises in measuring the degree of association between the correlated outcomes and the fixed effects. Here we describe an $R^2$ measure only for item I, i.e., comparing nested mean models with the same covariance structure. For our $R^2$ measure intended to evaluate fixed effects (mean differences), we specify a null model with only the intercept in the fixed effects. To compare nested mean models, we require the *same* covariance structure for both the null model and the model of interest. As a result of using the LRT in the GLMM to define a model $R^2$, we compare

the following two models

1. Model of Interest $\quad g[E(Y_{ij}|\boldsymbol{u}_i)] = \boldsymbol{x}'_{ij}\boldsymbol{\beta} + \boldsymbol{z}'_{ij}\boldsymbol{u}_i$
2. Null Model $\quad\quad\quad\quad g[E(Y_{ij}|\boldsymbol{u}_i)] = \quad \beta_0 + \boldsymbol{z}'_{ij}\boldsymbol{u}_i.$

As stated in Section 1, the LRT statistic has been used to define an $R^2$ measure for linear models with non-normal data, such as the multiple logistic regression model. The $R^2$ measure based on the LRT statistic for such models is given by

$$R_L^2 = 1 - \exp\left(-\frac{\text{LRT}}{N}\right) \tag{3}$$

where LRT is the likelihood ratio test statistic. Here $N$ is the number of independent sampling units.

We extend (3) to the GLMM with normal mixing distribution by substituting the LRT statistic for testing the appropriate set of model coefficients. The most common situation involves a model including an intercept and a hypothesis excluding the intercept, giving $H_0 : \boldsymbol{C}\boldsymbol{\beta} = \boldsymbol{0}$ for $\boldsymbol{C} = [\,\boldsymbol{0}_{(q-1)\times 1}\,\boldsymbol{I}_{q-1}\,]$ of rank $q-1$. For testing $H_0 : \boldsymbol{C}\boldsymbol{\beta} = \boldsymbol{0}$, the LRT statistic for the GLMM is given by

$$\text{LRT} = -2\left[l(\widetilde{\boldsymbol{\beta}}, \widetilde{\boldsymbol{\Sigma}}|\boldsymbol{Y}) - l(\widehat{\boldsymbol{\beta}}, \widehat{\boldsymbol{\Sigma}}|\boldsymbol{Y})\right], \tag{4}$$

where the LRT statistic contrasts the log-likelihood $l(\widetilde{\boldsymbol{\beta}}, \widetilde{\boldsymbol{\Sigma}}|\boldsymbol{Y})$ of the MLE $\widetilde{\boldsymbol{\beta}}$ obtained under the restricted model (null model) with the log-likelihood $l(\widehat{\boldsymbol{\beta}}, \widehat{\boldsymbol{\Sigma}}|\boldsymbol{Y})$ of the MLE $\widehat{\boldsymbol{\beta}}$ obtained under the unrestricted model (full model). The LRT statistic is asymptotically chi-square with $q-1$ degrees of freedom. See Tuerlinckx et al. (2006) for further details on the LRT for the GLMM.

Hence, under the previous assumptions, an $R^2$ measure for the GLMM can be given by

$$R_L^2 = 1 - \exp\left(\frac{\text{LRT}}{n}\right), \tag{5}$$

where $n = \sum_{i=1}^{N} p_i$, the total number of observations. By definition, $0 \leq R_L^2 \leq 1$, with $R_L^2 = 0$ indicating no multivariate association between $y$ and $X$. On the other hand, as $R_L^2$ nears 1, then the multivariate association between $y$ and $X$ becomes perfect.

In the linear univariate and multivariate models, adding a fixed effect can either explain more of the variance or add no additional explanation. Hence, for the linear univariate and multivariate models, the true population $R^2$ and estimated $R^2$ increase or remain the same. However, in the linear mixed model, adding a predictor in the fixed effects (between-subject) can increase the estimated variance of the random effects (within-subject effect) and hence increase the estimated variance of the response (Edwards et al. 2008) which may result in a decrease in $R_\beta^2$. This may also be the case for $R_L^2$ which requires further research. In such cases, $R_L^2$ is interpreted as indicating a decrease in measure of association possibly due to either misspecification of the "full" model and/or of sampling variation resulting in changes to the variance components estimates. However, the true population $R^2$ that $R_L^2$ estimates, under suitable conditions, should not decrease when a predictor is added.

## 4. Example Computations and Interpretations

Edwards et al. (2008) provided a linear mixed model example for a repeated blood pressure (BP) study (Fisher et al. 2008) that showed little association between repeated systolic and diastolic BP outcomes and a set of fixed effects. Using the same data, we construct a repeated dichotomous outcome (controlled or uncontrolled BP), fit a GLMM with the same fixed effects used for the linear mixed model, and compute $R_L^2$. Using the JNC VII classification of BP, BP is considered controlled if systolic BP is less than 140 mmHg and diastolic BP is less than 90 mmHg. We created a binary outcome that indicates whether a person's BP was controlled or uncontrolled at the time of measurement. Thus, for each subject, we have longitudinal binary data indicating controlled or uncontrolled. We then fit a GLMM with logit link and with random intercept and slope to this data to determine BP control over time.

Data from a retrospective longitudinal cohort study of 459 adults with hypertension (Fisher et al. 2008) illustrate the problem and the utility of our statistic. Longitudinal blood pressure (BP) level were taken on patients making at least four visits to the Family Practice Center at UNC during a two year period, 1999-2001. Predictor variables in the GLMM include indicators for Continuity of Care, and Race, Gender, Insurance status, Provider type, Marital status, as well as continuous linear Age at first measurement and linear Time. The random effects include intercept only with scalar $\Sigma$. Many studies have confirmed that blacks, on average, have higher BP than whites, and by proxy, less BP control. In this study, higher probability of uncontrolled BP was associated with blacks versus whites with p-value $= 0.0085$. For the linear mixed model analyses, higher BP was associated with blacks versus whites with for both systolic BP (p-value $= 0.0042$) and diastolic BP (p-value $= 0.0053$). The race effect remained the same across time (there was no Race $\times$ Time interaction).

For the fitted GLMM, the model $R_L^2 = 0.02$ with $n = 4768$ and $-2\left[l(\widetilde{\boldsymbol{\beta}}, \widetilde{\boldsymbol{\Sigma}}|\boldsymbol{Y}) - l(\widehat{\boldsymbol{\beta}}, \widehat{\boldsymbol{\Sigma}}|\boldsymbol{Y})\right] = -93.83$. For the linear mixed model analyses, $R_\beta^2 = 0.09$ for systolic BP and $R_\beta^2 = 0.27$ for diastolic BP. For both the GLMM and the linear mixed model, predictors were Continuity of Care, Gender, Insurance status (three levels and hence two dummy variables), Provider type, Marital status, linear Age, linear Time, Race. For the GLMM, the value of $R_L^2$ is consistent with the notion that for this study there is a very weak association between the repeated outcomes and the fixed effects.

## 5. Conclusions

In more general models there can be a large number of potential $R^2$'s from which to choose. This variety is due to the many interpretations that can be given to $R^2$ in the standard model. Each interpretation may motivate one or more possible measures (Magee 1990, Edwards et al. 2008). Given its underlying principles and due to the lack of available measures of association in the GLMM, $R_L^2$ should be used as a measure of association for fixed effects in the GLMM. Further research is required to explore the properties of $R_L^2$.

One drawback to $R_L^2$ is that both the log-likelihood must be computed for two model, both the model of interest and the null model. For the linear mixed model, only a single model is needed (Edwards et al. 2008). Also, readers familiar with the $R^2$ statistic for the linear univariate and multivariate models may at first be skeptical of a feature of $R_L^2$ that allows the measure to decrease when adding predictors. In the linear univariate and multivariate models, adding a fixed effect results in an increase (or no change) in the amount of variance explained by the predictors and hence the monotonic property of both the sample $R^2$ and true population $R^2$. However, similar to $R_\beta^2$ for the linear mixed model, $R_L^2$ may decrease by adding a predictor in the fixed effects. In such cases, $R_L^2$ is interpreted as indicating a decrease in measure of association possibly due to either misspecification of the "full" model and/or of sampling variation resulting in changes to the variance components estimates. However, the true population $R^2$ that $R_L^2$ estimates, under suitable conditions, should not decrease when a predictor is added.